\documentclass[12pt]{article}

\usepackage{amsmath} 
\usepackage{amssymb}
\usepackage{amsfonts}
\usepackage{amsthm}

\DeclareMathOperator{\Tr}{Tr}

\newtheorem{proposition}{Propostion}
\newtheorem{lemma}{Lemma}

\theoremstyle{definition}
\newtheorem*{demo}{Proof}
\newtheorem{definition}{Definition}

\begin{document}
	\begin{center}
		\Large
	\textbf{Pseudomode approach and vibronic non-Markovian phenomena in light harvesting complexes}\footnote{The research was supported by RSF (project No. «17-71-20154»).}	
	
		\large 
		\textbf{A.E. Teretenkov}\footnote{Steklov Mathematical Institute of Russian Academy of Sciences, Moscow, Russia;  Lomonosov Moscow State University, Moscow, Russia\\ E-mail:taemsu@mail.ru}
		\end{center}
		
			\footnotesize
			The pseudomode approach is discussed with an emphasis to  Gorini-Kossakowski-Sudarshan-–Lindblad form of this approach. The connection of the pseudomode approach with solutions of both the Friedrichs model and Jaynes-–Cummings model  with dissipation at zero temperature is shown. The obtained results are applied to non-Markovian phenomena description in the Fenna-Matthews-Olson complexes, estimations based on experimental data are done. A deformation approach to generalization of the pseudomode approach to the finite-temperature case is discussed.
			\normalsize

	\section{Introduction}
	We use the pseudomode approach developed in  \cite{Garraway96,Garraway97,Garraway97a,Dalton01,Garraway06} to describe non-Markovian evolution caused by strong coupling of a system and a reservoir. We apply the approach to the theory of non-Markovian exciton transport phenomena in light harvesting complexes. Vibrational degrees of freedom play the role of the reservoir in such systems. And states arising as a result of the strong coupling between excitons and these vibrational degrees of freedom are called vibrons. Thus, we refer to non-Markovian phenomena due to strong coupling as vibronic ones.
	
	The pseudomode approach allows one to describe the evolution of a system which is strongly coupled with a reservoir if the following conditions are met (more specific and strict statements could be found below in the main text of the article):
	\begin{enumerate}
		\item The system has a finite number of levels.
		\item The reservoir is at zero temperature.
		\item Interaction with the reservoir is described by a spectral density with a finite number of poles in the lower half-plane.
	\end{enumerate}
	Then this approach allows one to describe evolution of the system in terms of solutions for a finite number of linear ordinary differential equations. Similar to  \cite{Garraway97, Garraway97a} we are especially interested if it is possible to represent these equations as a master equation with a Gorini-Kossakowski-–Sudar\-shan-–Lindblad (GKSL) generator \cite{Gorini76, Lindblad76}.

	In contrast with \cite{Garraway96, Garraway97, Garraway97a, Dalton01, Garraway06} we separate three subjects, which, in our opinion, are logically independent, in the pseudomode approach. These three subjects correspond to three sections  \ref{sec:RecNorm}-\ref{sec:FriedMod} of our article. Such a presentation allows us to clarify the relation between the pseudomode approach and the solution of the Jaynes-–Cummings model  with dissipation at zero temperature \cite{Sachdev84} as well as the Friedrichs model \cite{Friedrichs48}.
	
	In section \ref{sec:RecNorm} we define a map which allows one to construct an equation with a GKSL generator by evolution with non-Hermitian Hamiltonian and non-increasing normalization. Let us note that such a map was discussed in \cite{Garraway97}. But GKSL equations obtained there were equations in the infinite-dimensional Hilbert space. Hence their solution is much more complex then the solution of initial equations with a non-Hermitian Hamiltonian. Moreover, we weaken conditions from \cite{Garraway97} on the non-Hermitian Hamiltonian.
	
	In section \ref{sec:oneExSecQuant} we start with the GKSL equation from the previous section and construct a GKSL equation in a larger (possibly infinite-dimensional) Hilbert space. We present the former GKSL equation which is a restriction of the latter one on 0-particle and 1-particle subspaces. Actually this is how the infinite-dimensional GKSL equations mentioned above arise in \cite{Garraway97}. They have a form of equations which describes the Jaynes-–Cummings model  with dissipation at zero temperature. However, the solution of these equations from \cite[formula (5.2)]{Sachdev84} is just based on the restriction of the initial model on 0-particle and 1-particle subspaces. The results of section \ref{sec:oneExSecQuant} allow us to generalize these solution on the case of multiple two-level atoms being strongly coupled with multiple bosonic modes.

	In section \ref{sec:FriedMod} we present the pseudomode approach just as the method to solve the Friedrichs model which reduces this model to evolution with non-Hermitian Hamiltonian. And only after that we use the results of sections \ref{sec:RecNorm} and \ref{sec:oneExSecQuant}  to show  how this approach allows one to solve a non-Markovian problem for the spin-boson model in the rotating wave approximation in terms of finite-dimensional GKSL equations, which was the original purpose to introduce this approach in \cite{Barnett86}. 
	
	In section \ref{sec:NonMark} we apply the pseudomode approach to describe non-Markovian decay of the two-level system in resonance with the frequency of the spectral density peak similar to paragraph 10.1.2 from \cite{Breuer10}. However, we describe the transition into Markovian limit in terms of Van Hove-Bogolyubov scaling \cite{Accardi2002} for the reduced density matrix rather than time-convolutionless projection operator method from \cite{Breuer10}.
	
	In section \ref{sec:expResults} we discuss the application of the obtained results to describe the non-Markovian phenomena in the Fenna-Matthews-Olson (FMO) light harvesting complexes at cryogenic tem\-pe\-ratures. We show that the presence of a strong coupling with a reservoir leads to long-living coherences and populations oscillations in non-degenerate systems. Such a behavior was discovered experimentally in \cite{Engel07}.
	
	In section \ref{sec:finTemp} we develop a deformation approach to description of non-Markovian phenomena at a finite temperature. Namely, we heuristically write a generator in the non-Markovian case. But then we show that the reduced density matrix in the Van Hove-Bogolyubov limit satisfies a Markovian master equation and thus one could regard our generator as a deformation of the Markovian one. Moreover, in the zero temperature limit our generator coincides with the exact generator from the previous section.
	
	In the conclusion section we summarize the main results of our work and discuss several  directions for further studies.
	
\section{Normalization reconstruction}\label{sec:RecNorm}

\begin{definition}\label{def:normRec}
	Let $ R $ be such a nuclear self-adjoint non-negatively definite operator in the Hilbert space $ \mathcal{H} $ that $ \Tr R \leqslant 1 $ (hereinafter we call this operator \textit{non-normalized density matrix}). Let us map this operator into the operator $ \rho $ in $ \mathcal{H} \oplus \mathbb{C} $ by formula
	\begin{equation}\label{eq:normRec}
	\rho = R \oplus 0 + (1-\Tr R)|0\rangle\langle 0 |,
	\end{equation}
	where $ |0 \rangle $ is the basis vector in $ \mathcal{H} \oplus \mathbb{C}  $ corresponding to the subspace $ \mathbb{C} $. We call such a map \textit{normalization reconstruction}.
\end{definition}

The operator $ \rho $  is obvious to be a density matrix (a self-adjoint non-negatively definite operator which trace equals one) in $ \mathcal{H} \oplus \mathbb{C}  $. Let us note that if the matrix $ R $ is already ''well-normalized'', i.e. $ \Tr R = 1 $, then, nevertheless, we obtain the new matrix $ \rho = R \oplus 0 $ in the ''larger'' Hilbert space $ \mathcal{H} \oplus \mathbb{C}  $ according to our definition. It could seem to be unnatural but we will see below that our definition is useful. If $  R $ is projector on the vector $ |\psi \rangle $ we define that the pure state $ |\psi \rangle \oplus 0 $ is a normalization reconstruction of the pure state $ |\psi \rangle $.

Let us consider a finite-dimensional Hilbert space $ \mathbb{C}^{n} $ and  a matrix $ H_{\rm eff} \in \mathbb{C}^{n\times n}$. Let us pose a Cauchy problem for the matrix-valued function $ R(t) : \mathbb{R}_+ \rightarrow \mathbb{C}^{n \times n}$
\begin{equation}\label{eq:nonHerEq}
\frac{d}{dt} R(t) = - i H_{\rm eff} R(t)  + i R(t)  H_{\rm eff}^{\dagger}, 
\end{equation}
\begin{equation}\label{eq:nonHerInit}
R(0) = R_0,
\end{equation}
where $ R_0 $ is a non-normalized density matrix and $ ^{\dagger} $ is a Hermitian conjugation.

\begin{definition}
	We call equation \eqref{eq:nonHerEq} \textit{a von Neumann equation with the non-Hermitian Hamiltonian} $ H_{\rm eff} $.
\end{definition}

\begin{definition}
	Let the solution $ R(t) $ of Cauchy problem \eqref{eq:nonHerEq}-\eqref{eq:nonHerInit} satisfy $ \Tr R(t) \leqslant 0  $ for
	arbitrary initial non-normalized density matrix $ R_0 $0, then we call
	equation \eqref{eq:nonHerEq} \textit{von Neumann equation with the non-Hermitian
		Hamiltonian $ H_{\rm eff} $ and non-increasing normalization}.
\end{definition}

\begin{lemma}\label{lem:reprOfHeff}
	Equation \eqref{eq:nonHerEq} is a von Neumann equation with a non-Hermitian
	Hamiltonian and non-increasing normalization if and only if $ H_{\rm eff} $ could be presented in the following form
	\begin{equation}\label{eq:reprOfHeff}
	H_{\rm eff} = H -\frac{i}{2}\sum_{l=1}^n \gamma_l |l \rangle \langle l |,
	\end{equation}
	 where $ H = H^{\dagger}$, $ | l \rangle , l=1, \ldots, n$ is a certain orthonormal basis in $ \mathbb{C}^n $: $ \langle l | l'\rangle = \delta_{ll'} $ and $ \gamma_l $ are non-negative real numbers: $ \gamma_l \geqslant 0 $.
\end{lemma}

\begin{demo}
	1) Let   \eqref{eq:nonHerEq} be a solution of the von Neumann equation with a non-Hermitian Hamiltonian and non-increasing normalization. Let us apply the trace to both sides of equation \eqref{eq:nonHerEq}
	\begin{equation*}
	\frac{d}{dt} \Tr R(t) = \Tr  \frac{H_{\rm eff} - H_{\rm eff}^{\dagger}}{i}R(t) ,
	\end{equation*}
	then $ \frac{d}{dt} \Tr R(t)|_{t=0} \leqslant 0 $ for any initial condition is equivalent to   $ \frac{H_{\rm eff} - H_{\rm eff}^{\dagger}}{i} \leqslant 0 $ as one could assume $ R(0) = |v\rangle \langle v|$ . Then non-negatively definite Hermitian matrix could be represented in the form $ \frac{H_{\rm eff} - H_{\rm eff}^{\dagger}}{i} = -\sum_{l} \gamma_l |l \rangle \langle l |$, where $ \gamma_l \geqslant 0 $ are opposite to eigenvalues of this matrix and $ |l \rangle $ are eigenvectors of the matrix forming an orthonormal basis. Besides, let us define $H =\frac{1}{2} (H_{\rm eff}+ H_{\rm eff}^{\dagger})$. Thus, we have \eqref{eq:reprOfHeff}.
	
	2) Inversely, if we assume \eqref{eq:reprOfHeff}, then it follows from \eqref{eq:nonHerEq} that
	\begin{equation}\label{eq:RtraceEv}
	\frac{d}{dt} \Tr R(t) = - \sum_l \gamma_l   \langle l | R(t) |l \rangle,
	\end{equation}
	which ensures non-increasing normalization for an arbitrary matrix $ R(t) \geqslant 0 $. The fact that the condition $ R(t) \geqslant 0 $  is conserved by equation \eqref{eq:nonHerEq} follows from the explicit solution $ R(t) = e^{- i  H_{\rm eff} t} R(0) e^{i  H_{\rm eff}^{\dagger} t} $. Indeed, if $ R(0) \geqslant 0$, then $  \langle v 
	| R(t) |v \rangle = \langle v' | R(0) | v' \rangle \geqslant 0$, where $ | v' \rangle  = e^{i  H_{\rm eff}^{\dagger} t}| v \rangle  $. \qed
\end{demo}

\begin{proposition}\label{prop:GKSL}
	Let $ R_t $ be a solution of Cauchy problem \eqref{eq:nonHerEq}-\eqref{eq:nonHerInit} for the von Neumann
	equation with the non-Hermitian Hamiltonian $ H_{\rm eff} $ and
	non-increasing normalization. Then the density matrix $ \rho_t $  obtained
	by normalization reconstruction \eqref{eq:normRec} is a solution of the Cauchy
	problem for the GKSL equation in
	the Lindblad form
	\begin{equation}\label{eq:GKSL}
	\frac{d}{dt} \rho(t) = - i [H \oplus 0, \rho(t)]+ \sum_l \left(L_l \rho(t) L_l^{\dagger} - \frac12 L_l^{\dagger} L_l \rho(t) - \frac12 \rho(t) L_l^{\dagger} L_l  \right),
	\end{equation} 
	\begin{equation*}
	\rho(0) = \rho_0,
	\end{equation*}
	where $ L_l = \sqrt{\gamma_l}|0 \rangle \langle l
	| $, $ \rho_0 $ is obtained by normalization
	reconstruction of $ R_0 $ and $ H $, $ \gamma_l $, $ |l \rangle  $ are defined in lemma \ref{lem:reprOfHeff}.
\end{proposition}

\begin{demo}
The matrix $ \rho(t) $ has form \eqref{eq:normRec} with respect to definition \ref{def:normRec}. Hence, we have
\begin{equation}\label{eq:derRho}
\frac{d}{dt} \rho(t) = \frac{d}{dt}R(t) \oplus 0 -\frac{d}{dt}\Tr R(t)|0\rangle\langle 0 |.
\end{equation}
Taking into account \eqref{eq:nonHerEq} and lemma \ref{lem:reprOfHeff} we obtain
\begin{align*}
\frac{d}{dt}R(t) \oplus 0 &= - i H_{\rm eff} R(t) \oplus 0 + i R(t)  H_{\rm eff}^{\dagger} \oplus 0 = \\
&=- i [H , R(t) ] \oplus 0 -\frac{1}{2}\sum_{l=1}^n \gamma_l |l \rangle \langle l | R(t) \oplus 0 -\frac{1}{2}   \sum_{l=1}^n \gamma_l R(t)|l \rangle \langle l |\oplus 0 = \\
&= - i [H \oplus 0, R(t) \oplus 0] -\frac{1}{2}\sum_{l=1}^n \gamma_l |l \rangle  \langle 0| | 0 \rangle\langle l | R(t) \oplus 0 -\frac{1}{2}   \sum_{l=1}^n \gamma_l R(t) \oplus 0  |l \rangle \langle 0| | 0 \rangle \langle l | =\\
&= - i [H \oplus 0, \rho(t)]  - \frac12 \sum_l \left(  L_l^{\dagger} L_l \rho(t) + \rho(t) L_l^{\dagger} L_l  \right),
\end{align*}
where the orthogonality condition $ \langle 0 |l \rangle =0, l \neq 0 $	and the definition $ L_l = \sqrt{\gamma_l}|0 \rangle \langle l
| $ are used.  Taking into account equation \eqref{eq:RtraceEv} we have
\begin{equation*}
\frac{d}{dt}\Tr R(t)|0\rangle\langle 0 | = - \sum_l \gamma_l   \langle l | R(t) |l \rangle |0\rangle\langle 0 | =  - \sum_l \gamma_l   |0\rangle \langle l | R(t) \oplus 0 |l \rangle \langle 0 | =  - \sum_l L_l \rho(t) L_l^{\dagger}.
\end{equation*}
By substituting these expression into \eqref{eq:derRho} we obtain \eqref{eq:GKSL}. \qed
\end{demo}
	
Just as a usual von Neumann equation in case of a pure initial state could be reduced to a Schroedinger equation, the solution of the Cauchy problem \eqref{eq:nonHerEq}-\eqref{eq:nonHerInit} with the initial condition $ R_0 = |\tilde{\psi}_0\rangle \langle \tilde{\psi}_0| $ could be represented in the form $ R(t) = |\tilde{\psi}(t)\rangle \langle \tilde{\psi}(t)| $, where $ |\tilde{\psi}(t)\rangle $ is a solution of the equation
\begin{equation}\label{eq:ShroEq}
\frac{d}{dt}|\tilde{\psi}(t)\rangle = - i H_{\rm eff}|\tilde{\psi}(t)\rangle,
\end{equation} 
with the initial condition $ |\tilde{\psi}(0)\rangle = |\tilde{\psi}_0\rangle$. 

\begin{definition}
	Let us call equation \eqref{eq:ShroEq}, where $ H_{\rm eff} $ is defined by formula \eqref{eq:reprOfHeff}, \textit{a Schroedinger equation with the non-Hermitian Hamiltonian and non-increasing normalization}.
\end{definition}

It is equation \eqref{eq:ShroEq} that was considered in \cite{Garraway97}. In this section we have introduced its natural generalization \eqref{eq:nonHerEq}.

\section{One-particle second quantization}\label{sec:oneExSecQuant}

Let us consider a Hilbert space of the form $ \mathbb{C}^{n} \oplus \mathbb{C} $. Let $ \mathcal{X}_i $ be auxiliary Hilbert spaces with $ \dim \mathcal{X}_i \geqslant 2$. Let us introduce the linear injection $\hat{} : \mathbb{C}^{n} \oplus \mathbb{C} \rightarrow \otimes_{i=1}^{n} \mathcal{X}_i $ which is defined on the basis by the following rule
\begin{equation}\label{eq:OneExSecQuant}
\begin{aligned}
| l \rangle  &\rightarrow | \hat{l} \rangle  = | 0 \rangle_1 \otimes \cdots \otimes| 0 \rangle_{l-1} \otimes |1 \rangle_{l} \otimes | 0 \rangle_{l+1} \otimes \cdots \otimes | 0 \rangle_{n} , \qquad l \neq 0; \\
| 0 \rangle  &\rightarrow | \hat{0} \rangle  = | 0 \rangle_1  \otimes \cdots \otimes | 0 \rangle_{n} .
\end{aligned}
\end{equation}

We call such a map \textit{a one-particle second quantization}. The term ''one-particle'' refers to the fact that we consider only the states in $  \otimes_{i=1}^{n} \mathcal{X}_i $  which correspond to the image of  $ \mathbb{C}^{n} \oplus \mathbb{C} $ for such an injective map. In a similar way let us map  the density matrices $  \rho = \sum_{l,k =0}^n \rho_{lk}|l\rangle \langle k| $ into the matrices
\begin{equation}\label{eq:OneExSecQuantDenMat}
 \hat{\rho} = \sum_{l,k =0}^n \rho_{lk}|\hat{l}\rangle \langle \hat{k}|.
\end{equation}

\begin{proposition}\label{prop:OneExSecQuantGKSL}
	Let $ a_l $ be arbitrary operators such that $ a_l^{\dagger} | \hat{0} \rangle = | \hat{l} \rangle $ and  $ a_l | \hat{l} \rangle = |\hat{0} \rangle $. Let $ \rho_t $ satisfy equation \eqref{eq:GKSL} with coefficients defined in
	proposition \ref{prop:GKSL}. Then $ \hat{\rho}_t $ defined by formula \eqref{eq:OneExSecQuantDenMat} satisfies the equation
	\begin{equation}\label{eq:OneExSecQuantGKSL}
	\frac{d}{dt} \hat{\rho}(t) = - i [\hat{H}, \hat{\rho}(t)]+ \sum_{l=1}^n \gamma_l \left(a_l \hat{\rho}(t) a_l^{\dagger} - \frac12 a_l^{\dagger} a_l \hat{\rho}(t) - \frac12 \hat{\rho}(t) a_l^{\dagger} a_l  \right), \qquad \hat{H} = \sum_{l,k = 1}^n H_{lk} a_l^{\dagger} a_k.
	\end{equation} 

\end{proposition}
The proof of this proposition is based on direct substitution.

There are three usual choices for operations $ a_l $ in the physical applications: bosonic annihilation operators $ b_l $ satisfying canonical commutation relations $ [b_l, b_k^{\dagger}] = \delta_{lk}, [b_l, b_k] = 0 $, fermionic anni\-hi\-la\-tion operators $ f_l $ satisfying canonical anticommutation relations $ \{f_l, f_k^{\dagger}\} = \delta_{lk},  \{f_l, f_k\} = 0 $ or two-level annihilation operators $ \sigma_l $ satisfying $ \{\sigma_l, \sigma_l^{\dagger}\} = I,  \{\sigma_l, \sigma_l\} = 0 , [\sigma_l, \sigma_k] =0,  [\sigma_l, \sigma_k^{\dagger}] =0,  l \neq k $. It is important to note that it is not necessary to choose the operators of the same type for all the indices. For example, one could choose some of $ a_l $ to be bosonic and some of them to be two-level. This is the case of interaction of two-level atoms and modes of a bosonic (electromagnetic, phonon etc.) field. The operators of different kind are assumed to commute with each other.

In particular, both the Jaynes-–Cummings model (introduced in \cite{Jaynes63},  bibliographical review could be found in \cite{Shore93}) and its dissipative zero-temperature generalization \cite{Sachdev84, Barnett86, Puri86, Wonderen97} have form \eqref{eq:OneExSecQuantGKSL}. Namely, the Jaynes-–Cummings model with dissipation at zero temperature is defined by the equation
\begin{equation*}
\frac{d}{dt} \hat{\rho}(t) = - i [\omega_1 \sigma^{\dagger} \sigma + \omega_2 b^{\dagger} b + g (\sigma^{\dagger} b + \sigma b^{\dagger}), \hat{\rho}(t)]+ \gamma \left(b \hat{\rho}(t) b^{\dagger} - \frac12 b^{\dagger} b \hat{\rho}(t) - \frac12 \hat{\rho}(t) b^{\dagger} b  \right).
\end{equation*}
Proposition \ref{prop:OneExSecQuantGKSL} allows one to reduce the problem of finding one-particle solutions of this equation to the solution of the following equation:
\begin{align*}
\frac{d}{dt} \rho(t) = - i [\omega_1 |1 \rangle \langle 1| + \omega_2 |2 \rangle \langle 2| + g (|1 \rangle \langle 2| + |2 \rangle \langle 1|), \rho(t)]+ \\
+\gamma \left(|0 \rangle \langle 2| \rho(t) |2 \rangle \langle 0| - \frac12 |2 \rangle \langle 2| \rho(t) - \frac12 \rho(t) |2 \rangle \langle 2| \right)
\end{align*}
which has form \eqref{eq:OneExSecQuantGKSL}.

Let us also note that if we assume all the operators $ a_l $ in equation \eqref{eq:OneExSecQuantGKSL} to be either only bosonic ones or only fermionic ones then we obtain a particular case of the GKSL equation with a quadratic generator \cite{Dodonov88, Heinosaari09, Teret16, Teret17, Teret17f}. Gaussian solutions could be obtained for such equations. However, in this case one could also obtain one-particle solutions as an example of the simplest non-Gaussian solutions for GKSL editions with quadratic generators.

The most important feature of the map \eqref{eq:OneExSecQuant} is that it allows one to consider the initial system as a composite one. In particular, one could calculate partial traces with respect to subspace $ \mathcal{X}_i $, where $ i \in I $ and $ I $ is a subset of the set of numbers from $ 1 $ to $ n $. Moreover, if the density matrix is obtained by formula \eqref{eq:OneExSecQuantDenMat}, then the particular trace has very simple expression:

\begin{proposition}\label{prop:trace}
	Let $ \hat{\rho} = \sum_{l,k =0}^n \rho_{lk}|\hat{l}\rangle \langle \hat{k}| $, then the partial trace with respect to spaces indexed by $ I $ could be calculated by the formula:
	\begin{equation}\label{eq:trace}
	\Tr_{ \mathcal{X}_i, i \in I} \hat{\rho} = \hat{\rho}_{\overline{I}} + \left(\rho_{00} + \sum_{l \in I} \rho_{ll}\right) | \hat{0} \rangle \langle \hat{0} |, \quad \hat{\rho}_{\overline{I}} = \sum_{l,k \in \overline{I}} \rho_{lk}|\hat{l}\rangle \langle \hat{k}|,
	\end{equation}
	where $  \overline{I} = \{0, \ldots, n\} \setminus I$.
\end{proposition}
One could prove this proposition by direct calculation. Let us emphasize that according to our definitions $ 0 $ is always contained in $ \overline{I} $ rather than $ I $.

Thus, if one traces out the subspaces indexed by $ I $, then the result is just a matrix without entries corresponding to the indices from $ I $ and with vacuum state population changed to its initial value plus the sum of populations corresponding to the indices from $ I $. The latter one guarantees normalization conservation.

Moreover, this proposition motivates the concept of the trace with respect to a set of indices for initial $ (n +1) \times (n+1)$ matrices  $ \rho $ by formula
\begin{equation}\label{eq:traceI}
\Tr_I \rho = \rho_{\overline{I}} + \left(\rho_{00} + \sum_{l \in I} \rho_{ll}\right) | 0 \rangle \langle 0 |, \quad \rho_{\overline{I}} = \sum_{l,k\in \overline{I}} \rho_{lk }|l\rangle \langle k|.
\end{equation}

In general, the embedding of the density matrices of so-called non-composite systems  (see \cite{Chernega14a}) into spaces which have a tensor product form appeared to be quite fruitful for the derivation of an entropy inequalities. The non-composite system is a system described by a Hilbert space which is not of the tensor product form or at least this form is not given explicitly. General discussion (far beyond our particular one-particle case) and more complete bibliography on this topic could by found in \cite{Chernega14, Chernega13, Manko18}. In regard to light harvesting complexes the trace with respect to an index set was discussed in \cite{Bradler10} for mutual information calculation. The results of this work were discussed in \cite{Arefeva16} from the holography point of view.

\section{Friedrichs model and pseudomode approach}\label{sec:FriedMod}

The Friedrichs model was introduced in \cite{Friedrichs48} and has been widely discussed in scientific papers in connection with the perturbation theory for continuous spectrum \cite{Faddeev64}, non-integrate large Poincare's systems and spontaneous time-symmetry breaking \cite{Petrosky91, Prigogine95, Antoniou93, Karpov00}, Gamow vectors in the rigged Hilbert spaces \cite{Parravicini80, Ordonez04, Antoniou03}. Moreover, this model has generalizations which allow one to obtain non-Markovian equations which are second order differential equations with respect to time \cite{Kossakowski07}. We rather focus on the features of the Friedrichs model which are related to pseudomode theory developed in \cite{Garraway96, Garraway97, Garraway97a, Dalton01, Garraway06}.

The Friedrichs model is defined by the Hamiltonian in $ \mathbb{C} \oplus L^2(\mathbb{R})$:
\begin{equation}\label{eq:Ham}
H_{F} = \int \omega_k |k \rangle  \langle k|  d k + \omega_1 |1 \rangle \langle 1| +  \int \left(  g_{ k}^*  | k \rangle \langle 1 | +   g_{k}  | 1 \rangle \langle k | \right) d k.
\end{equation}

 The spin-boson model in the rotating wave approximation is one of possible second quantizations according to proposition \ref{prop:OneExSecQuantGKSL}. It is defined by the Hamiltonian in $ \mathbb{C}^2 \otimes \mathfrak{F}_b(L^2(\mathbb{R})) $
\begin{equation}\label{eq:SBHam}
\hat{H}_{SB} = \int \omega_k I \otimes b_k^{\dagger} b_k d k + \omega_1 |1 \rangle \langle 1| \otimes I +  \int \left(  g_{ k}^*  | 0 \rangle \langle 1 | \otimes b_k^{\dagger}+   g_{k}  | 1 \rangle \langle 0 | \otimes b_k \right) d k.
\end{equation}

This is the model being solved in \cite{Garraway96, Garraway97, Garraway97a, Dalton01, Garraway06}. However, as the temperature of the reservoir was assumed equal to zero, then the one-particle solutions of the Schroedinger equation with Hamiltonian \eqref{eq:SBHam} were considered.  Hence, the problem was actually reduced to the solution of the Schroedinger equation with the Hamiltonian $ H_{F} $, even though it was not mentioned in these works.

Let us consider the Schroedinger equation with such a Hamiltonian
\begin{equation}\label{eq:Shr}
\frac{d}{dt}|\psi(t)\rangle =- i H_F |\psi(t)\rangle,
\end{equation}
with the initial condition
\begin{equation}\label{eq:initCond}
|\psi(0)\rangle  =  \psi_1(0) |1\rangle,
\end{equation}
then the following proposition is held.

\begin{proposition}
	The Cauchy problem \eqref{eq:Ham}, \eqref{eq:Shr}, \eqref{eq:initCond} has the solution
	\begin{equation}\label{oneExSol}
	|\psi(t)\rangle  =  \psi_1(t) |1\rangle + \int dk\psi_k(t)  |k \rangle, 
	\end{equation}
	where  $ \psi_{1}(t) $ satisfies the equation
	\begin{equation}\label{eq:sysEvol}
	\frac{d \psi_1(t)}{dt} = - i \omega_1 \psi_1(t) - \int_0^t dt' G( t - t') \psi_1(t'),
	\end{equation}
	where $ G(t) = \int d k e^{-i \omega_k  t} |g_{k}|^2  $ and  $ \psi_k(t) $ could be expressed it terms of $ \psi_{1}(t) $ by the formula
	\begin{equation}\label{eq:psi_k}
	\psi_k(t)= - i  g_{k}^* \int_0^t d\tau  e^{i  \omega_k (\tau - t)} \psi_1 (\tau).
	\end{equation}
\end{proposition}

Actually the proof of this proposition could be found in \cite{Garraway97}. The main idea consists in the substitution of \eqref{oneExSol} into \eqref{eq:Shr}. As a result, we obtain the following equations
\begin{align*}
\frac{d}{dt} \psi_1(t) &=-i \omega_1 \psi_1(t) - i \int dk  g_{k} \psi_k(t),\\
\frac{d}{dt} \psi_k(t) &= - i  \omega_k \psi_k(t) - i  g_{k}^* \psi_1(t).
\end{align*}
We integrate the latter equation with initial condition \eqref{eq:initCond}, i.e. $ \psi_k(0) = 0 $, and obtain \eqref{eq:psi_k}. Then we substitute \eqref{eq:psi_k} into the former one and obtain \eqref{eq:sysEvol}.

The spectral density $ J(\omega) $ is usually defined in physical applications rather than the function $ G(t) $. These functions are related by the Fourier transform:
\begin{equation*}
G(t) = \int_{-\infty}^{+\infty} \frac{d \omega }{2 \pi}e^{-i \omega  t} J(\omega).
\end{equation*}

The psedomode approach is essentially a method for solving integro-differential equation \eqref{eq:Ham} in the case when $ G(t) $ is a finite linear span of exponents:
\begin{equation}\label{eq:goodG}
G(t) = \sum_{l=1}^n g_l^2 e^{-\left(\frac{\gamma_l }{2}+ i \omega_l\right) t }, \quad t>0,
\end{equation}
where $ \gamma_l >0 $, $ \omega_l \in \mathbb{R} $ and $ g_l $ are complex in the general case. The spectral densities with a finite number of poles in the lower half-plane lead to such a form of the function $ G(t) $ (see \cite{Garraway97}). In particular, the spectral density
\begin{equation*}
J(\omega) = \sum_{l=1}^n \frac{\gamma_l g_l^2}{\left(\frac{\gamma_l}{2}\right)^2+(\omega- \omega_l)^2}
\end{equation*}
leads to \eqref{eq:goodG}, i.e. the spectral density has the form of a sum of a finite number of peaks.

\begin{proposition}	
	Let  $ \psi_{1}(t) $ be a solution of equation \eqref{eq:sysEvol} with $ G(t) $ of form \eqref{eq:goodG}. Let us consider the Hilbert space $ \mathbb{C}^{n + 1} $ with basis $ |1 \rangle $, $ | \tilde{l} \rangle $. Let us define the non-Hermitian Hamiltonian $ H_{\rm eff} \in \mathbb{C}^{(n+1) \times (n+1)} $ by the formula
	\begin{equation}\label{eq:Heff}
	H_{\rm eff} = \omega_1 |1 \rangle \langle 1 |+\sum_{l=1}^n\left( \left(\omega_l - i\frac{\gamma_l}{2}\right)| \tilde{l} \rangle \langle \tilde{l} | + g_l ( |\tilde{l} \rangle \langle 1 | + |1 \rangle \langle \tilde{l} |) \right)
	\end{equation}
	then the vector
	\begin{equation}\label{eq:psi_tilde}
	|\tilde{\psi}(t) \rangle =  \psi_{1}(t)|1 \rangle  +  \sum_l \varphi_l(t)  |\tilde{l} \rangle,
	\end{equation}
	where 
	\begin{equation}\label{eq:pseudoPhi}
	\varphi_l(t) = -i g_l \int_0^t d\tau e^{-\left(\frac{\gamma_l }{2}+ i \omega_l\right)  (t-\tau) } \psi_1 (\tau),
	\end{equation}
	satisfies the equation
	\begin{equation}\label{eq:psiHeff}
	\frac{d}{dt} | \tilde{\psi}(t) \rangle = - i H_{\rm eff} | \tilde{\psi}(t) \rangle.
	\end{equation}
\end{proposition}

\begin{demo}
Taking into account definition  \eqref{eq:pseudoPhi} we represent equation \eqref{eq:sysEvol} in the following form
\begin{equation*}
\frac{d}{dt} \psi_1(t) =-i \omega_1 \psi_1(t)  - i\sum_l g_l \varphi_l(t).
\end{equation*}
Differentiating \eqref{eq:pseudoPhi} we have
\begin{equation*}
\frac{d}{dt} \varphi_l(t) = -i g_l \psi_1(t) - \sum_{l} \left(\frac{\gamma_l }{2}+ i \omega_l\right)  \varphi_l(t).
\end{equation*}
The obtained equations coincide with equation \eqref{eq:psiHeff} in the component-wise representation. \qed 
\end{demo}

In work \cite{Garraway97} it was discussed if the equation with non-Hermitian Hamiltonian of form \eqref{eq:Heff} can be represented in a GKSL form. However, the general method for such a representation was obtained only for the real $ g_l $ case. At the same time, proposition \ref{prop:GKSL} allows one to do it in the more general case when the matrix $ -i(H_{\rm eff} - H_{\rm eff}^*) $ is non-positive definite. Nevertheless, GKSL form \eqref{eq:GKSL} could be defined simpler if  $ g_l $ are real, as  $ -i(H_{\rm eff} - H_{\rm eff}^*) $ has already been diagonalized and, hence,  $ L_l = \sqrt{\gamma_l}|0 \rangle \langle l|  $ and $ 	H = \omega_1 |1 \rangle \langle 1 |+\sum_{l=1}^n\left( \omega_l | \tilde{l} \rangle \langle \tilde{l} | + g_l ( |\tilde{l} \rangle \langle 1 | + |1 \rangle \langle \tilde{l} |) \right) $.  

Similar to proposition \ref{prop:OneExSecQuantGKSL} the solution of the Schroedinger equation with Hamiltonian \eqref{eq:SBHam} with the initial condition $ |\hat{\psi}(0)\rangle  =  \psi_0(0) |1\rangle + \psi_1(0) |1\rangle $ has the form $ |\hat{\psi}(0)\rangle  = \psi_0(0) |0\rangle \otimes |0\rangle+ \psi_1(t) |1\rangle \otimes |0\rangle  + \int dk\psi_k(t)  |0 \rangle \otimes b_k^{\dagger} |0 \rangle $, where $ \psi_1(t)  $ and $ \psi_k(t) $ are defined by \eqref{eq:sysEvol} and \eqref{eq:psi_k}. According to proposition \ref{prop:trace} we calculate the reduced density matrix $ \hat{\rho}_S(t) \equiv \Tr_{\mathfrak{F}_b(L^2(\mathbb{R}))} |\hat{\psi}(t) \rangle \langle \hat{\psi} (t) |  $ by formula \eqref{eq:trace}:
\begin{equation*}
\hat{\rho}_S (t)  =  |\psi_{1}(t)|^2 |1 \rangle \langle 1 | + \psi_{1}(t) \psi_{0}^*(0) |1 \rangle \langle 0 | + \psi_{0}(0) \psi_{1}^*(t) |0 \rangle \langle 1 | + (1 - |\psi_{1}(t)|^2 )|0 \rangle \langle 0 | .
\end{equation*}
If the density matrix $ \rho $  (which is defined by \eqref{eq:normRec} and satisfies GKSL equation \eqref{eq:GKSL}) is a result of the normalization recovery map for the non-normalized density matrix $ |\tilde{\psi}(t) \rangle \langle \tilde{\psi}(t)| $, where $ |\tilde{\psi}(t) \rangle$ is defined formula \eqref{eq:psi_tilde}, then the partial trace of $ \rho $ with respect to indices $ \tilde{l} $ takes the form
\begin{equation*}
\rho_S (t)  =  |\psi_{1}(t)|^2 |1 \rangle \langle 1 | + \psi_{1}(t) \psi_{0}^*(0) |1 \rangle \langle 0 | + \psi_{0}(0) \psi_{1}^*(t) |0 \rangle \langle 1 | + (1 - |\psi_{1}(t)|^2 )|0 \rangle \langle 0 | = \hat{\rho}_S (t)
\end{equation*}
by formula \eqref{eq:traceI}.

Thus, we consider the spin-boson model in the rotating wave approximation with the reservoir which is initially in the vacuum state (at zero temperature) and $ G(t) $ has form \eqref{eq:goodG}. Then the reduced density matrix for this model is equal to the reduced density matrix $ \hat{\rho}_S (t) $, i.e. the trace of the matrix $ \rho(t) $ with respect to the indices, where $ \rho(t) $ is the solution of GKSL equation \eqref{eq:GKSL}. Hence, the states  $ |\tilde{l} \rangle $ were called pseudomodes in \cite{Garraway96}, because the unitary evolution of the system and the reservoir with infinite-number degrees of freedom is equivalent to GKSL evolution in the finite-dimensional Hilbert space from the point of view of reduced density matrix.

\section{Non-Markovian resonance decay}\label{sec:NonMark}

Let us consider the case when there is only one term in formula \eqref{eq:goodG}, i.e. $ G(t) = g^2 e^{-\left(\frac{\gamma }{2}+ i \omega_1\right) t } $. Moreover, we assume $ g $ to be real and the resonance condition $ \omega_0 = \omega_1 $ to be satisfied. Then formula \eqref{eq:Heff} takes the form
\begin{equation*}
H_{\rm eff} = \omega_1 | 1 \rangle \langle 1| + \left(\omega_1 - i  \frac{\gamma}{2} \right) | \tilde{1} \rangle \langle \tilde{1}| + g( |\tilde{1}\rangle \langle 1| + | 1 \rangle \langle \tilde{1}|).
\end{equation*}

As the Markovian limit is considered below and it is usually expressed in the interaction representation, so we also transform our Hamiltonian into the interaction representation
\begin{equation*}
H_{\rm eff}' \equiv e^{i \omega_1 (| 1 \rangle \langle 1| + | \tilde{1} \rangle \langle \tilde{1}|) t}(H_{\rm eff} - \omega_1 (| 1 \rangle \langle 1| + | \tilde{1} \rangle \langle \tilde{1}|)) e^{-i \omega_1 (| 1 \rangle \langle 1| + | \tilde{1} \rangle \langle \tilde{1}|) t} = - i  \frac{\gamma}{2}| \tilde{1} \rangle \langle \tilde{1}| + g( |\tilde{1}\rangle \langle 1| + | 1 \rangle \langle \tilde{1}|) ,
\end{equation*}
and we write equation \eqref{eq:psiHeff} in the interaction representation:
\begin{equation*}
\frac{d}{dt} \begin{pmatrix}
\psi_{1}(t)\\
\psi_{\tilde{1}}(t)
\end{pmatrix}
=\begin{pmatrix}
0 & - ig\\
- i g & - \frac{\gamma}{2}
\end{pmatrix}
\begin{pmatrix}
\psi_{1}(t)\\
\psi_{\tilde{1}}(t)
\end{pmatrix}.
\end{equation*} 
The solution in the case of initial condition $ \psi_{1}(0) =1, \psi_{\tilde{1}}(0) = 0 $ has the form
\begin{align*}
\psi_{1}(t)  &= e^{- \frac{\gamma}{4} t}  \left( \cosh \Delta t + \frac{\gamma}{4 \Delta} \sinh \Delta t \right),\\
\psi_{\tilde{1}}(t) &= - \frac{i g}{\Delta} e^{- \frac{\gamma}{4} t} \sinh \Delta t,
\end{align*}
where $ \Delta = \frac{1}{4}\sqrt{\gamma^2 - 16 g^2} $. The solution after taking the trace:
\begin{equation*}
\rho_S(t) = e^{- \frac{\gamma}{2} t}  \left( \cosh \Delta t + \frac{\gamma}{4 \Delta} \sinh \Delta t \right)^2 | 1 \rangle \langle 1| + \left( 1 -e^{- \frac{\gamma}{2} t}  \left( \cosh \Delta t + \frac{\gamma}{4 \Delta} \sinh \Delta t \right)^2 \right)| 0 \rangle \langle 0| .
\end{equation*}

In our last paragraph we have mainly reproduced in our notation the results of paragraph 10.1.2 from \cite{Breuer10}. However, the transition to the Markovian limit was expressed there in terms of time-dependent population transfer rate which occurs in the time-convolutionless projection operator method \cite{Shibata77, Breuer99}, \cite[Section 9.2]{Breuer10}. But we use Van Hove-Bogolyubov scaling as it was done in the stochastic limit (see \cite{Accardi2002}, particularly the discussion in sections 1.8 and 1.27). Let us consider this solution as a function of time and the coupling constant $ \rho_S(t, g) $. Then the transition to the Markovian limit is described by the formula 
\begin{equation*}
\lim\limits_{\lambda \rightarrow 0}\rho_{S}\left(\frac{1}{\lambda^2} t, \lambda g\right) = e^{- \frac{4 g^2 }{\gamma} t} | 1 \rangle \langle 1|  +(1- e^{- \frac{4 g^2 }{\gamma} t}) | 0 \rangle \langle 0|.
\end{equation*}
Thus, the decay rate in the Markovian limit $ \gamma_0 = \frac{4 g^2}{\gamma} $.

In the strong coupling limit the oscillating dynamics occurs
\begin{equation*}
\lim\limits_{\lambda \rightarrow + \infty}\rho_{S}\left(\frac{1}{\lambda} t, \lambda g\right) = \cos^2 ( g t) | 1 \rangle \langle 1| +  \sin^2 ( g t) | 0 \rangle \langle 0| .
\end{equation*}
Thus, the transition from the Markovian regime to the strongly non-Markovian one corresponds to the transition from the relaxation regime to the oscillation one.

Proposition \ref{prop:GKSL} allows us to write GKSL equation \eqref{eq:GKSL} for the density matrix $ \rho_t $ with the coefficients
\begin{equation*}
H = g( |\tilde{1}\rangle \langle 1| + | 1 \rangle \langle \tilde{1}|) , \qquad  L_1 = \sqrt{\gamma}|0 \rangle \langle 1|.
\end{equation*}
Namely, we have the equation
\begin{equation*}
\frac{d}{dt} \rho(t) = - i g [ |\tilde{1}\rangle \langle 1| + | 1 \rangle \langle \tilde{1}|, \rho(t)] + \gamma \left(|0 \rangle \langle 2| \rho(t) |2 \rangle \langle 0| - \frac12 |2 \rangle \langle 2| \rho(t)- \frac12 \rho(t)|2 \rangle \langle 2| \right).
\end{equation*}
Let us define $ D_{lk}(\rho) = |k \rangle \langle l| \rho |l \rangle \langle k| - \frac12 |l \rangle \langle l| \rho- \frac12 \rho|l \rangle \langle l| $, $ h_{kl}(\rho) = -i [|k\rangle \langle l| + | l \rangle \langle k|,\rho]$. This equation in terms of Markovian decay rate $ \gamma_0 $ takes the form:
\begin{equation*}
\frac{d}{dt} \rho(t) = \frac{4 g^2}{\gamma_0} D_{\tilde{1} 0} (\rho(t) ) + g h_{1 \tilde{1}}(\rho(t) ).
\end{equation*}

\begin{proposition}\label{prop:finTempMarkLim}
	Let $ \rho(t, g , \gamma_0) $ be a solution of the Cauchy problem 
	\begin{equation*}
	\frac{d}{dt} \rho(t, g , \gamma_0) = \frac{4 g^2}{\gamma_0} D_{\tilde{1} 0} (\rho(t, g , \gamma_0) ) + g h_{1 \tilde{1}}(\rho(t, g , \gamma_0) )
	\end{equation*}
	with the initial condition
	\begin{equation*}
	\rho(0, g , \gamma_0) = \rho_{11} | 1 \rangle \langle 1| + (1 - \rho_{11})| 0 \rangle \langle 0| + \rho_{10} | 1 \rangle \langle 0| + \rho_{01} | 0 \rangle \langle 1|.
	\end{equation*}
	Then  $ \rho_M(t, \gamma_0) \equiv \lim\limits_{\lambda \rightarrow 0} \rho_S \left(\frac{t}{\lambda^2}, \lambda g , \lambda^2 \gamma_0 \right) $ satisfies the equation
	\begin{equation*}
	\frac{d}{dt} \rho_M(t, \gamma_0) = \gamma_0 D_{1 0} (\rho_M(t, \gamma_0)).
	\end{equation*} 
\end{proposition}
The proof of this proposition is based on direct calculation of the reduced density matrix which takes form
\begin{align*}
\rho_S \left(t, g, \gamma_0 \right) =& \rho_{11}e^{- \frac{4 g^2}{\gamma_0}  t}  \left( \cosh \Delta t + \frac{g^2}{\gamma_0 \Delta}  \sinh \Delta t \right)^2 | 1 \rangle \langle 1| +\\
&+ \left( 1 - \rho_{11} e^{- \frac{4 g^2}{\gamma_0} t}  \left( \cosh \Delta t + \frac{g^2}{\gamma_0 \Delta}  \sinh \Delta t \right)^2 \right)| 0 \rangle \langle 0| +\\
&+ e^{- \frac{2 g^2}{\gamma_0}t}  \left( \cosh \Delta t + \frac{g^2}{\gamma_0 \Delta} \sinh \Delta t \right)(\rho_{10} | 1 \rangle \langle 0| + \rho_{01} | 0 \rangle \langle 1|), \qquad \Delta = g \sqrt{\left(\frac{g}{\gamma_0}\right)^2-1}
\end{align*}
Then, as $ \lambda \rightarrow 0 $, the limit takes the form
\begin{equation*}
\rho_M(t, \gamma_0) \equiv \lim\limits_{\lambda \rightarrow 0} \rho_S \left(\frac{t}{\lambda^2}, \lambda g , \lambda^2 \gamma_0 \right)  = \rho_{11} e^{- \gamma_0 t} | 1 \rangle \langle 1|  +(1- \rho_{11} e^{- \gamma_0 t}) | 0 \rangle \langle 0| + e^{-\frac{\gamma_0}{2} t} (\rho_{01}  | 0 \rangle \langle 1| + \rho_{10}  | 1 \rangle \langle 0|).
\end{equation*}
Let us note that besides the time-convolutionless projection operator method mentioned above the non-Markovian resonance decay was studied by means of master equations with a time-non-local generator in \cite{Breuer06}.

\section{Non-Markovian phenomena in FMO complexes}\label{sec:expResults}

The following features were discovered for FMO complexes at cryogenic temperature $ \beta^{-1} = 77 \, \text{K} $:
\begin{enumerate}
	\item Long-living coherences (about 1 ps), while the Markovian models predicted coherence lifetime not exceeding 250 fs \cite{Engel07}.
	\item Oscillation of the populations in the case of absence of degeneracy, which also does not coincide with the Markovian dynamics (see the discussion in \cite[p. 154]{Engel14}).
\end{enumerate}

These discoveries gave rise to the sufficient interest in the quantum phenomena in photosynthetic systems \cite{Collini10, Scholes11, Lee07, Mohseni08, Plenio08}. 

In spite of the fact that realistic models consider the FMO complex in the one-exciton ap\-pro\-xi\-ma\-tion (see \cite{Rebentrost09, Bradler10} for the details) as a 7-level system \cite{Bradler10}, we follow  \cite{Chin13, Plenio13} and consider an essentially simplified dimer model of the FMO complex. We also consider the one-exciton ap\-pro\-xi\-mation and assume the states $ |1 \rangle $ and $ |0 \rangle $ from the previous section to correspond to the so-called global basis (see the discussion in \cite{Trushechkin16}). 

Data from \cite{Plenio13} suggest $ \omega_0 = 202 \, \text{cm}^{-1}, \beta^{-1} \approx 53 \, \text{cm}^{-1}  $, then  $ n = \frac{1}{e^{\beta \omega_0}-1} \approx 0.02$, hence one could use zero-temperature approximation $ n \approx 0 $.  Moreover, if we assume the Huang-Rhys factor $ S = 0.02 $  as in \cite{Plenio13}, then the coupling constant $ g = \sqrt{S} \omega_0 \approx 29 \, \text{cm}^{-1}$. The typical transition time \cite{Engel07} predicted by the Markovian models equals 250 fs, which corresponds the Markovian coherence decay rate $  \frac{\gamma_0}{2} = 133 \,\text{cm}^{-1}$ . Let us evaluate $ \frac{\gamma}{4}  = g^2/\gamma_0 \approx 3.2 \,\text{cm}^{-1}$, i.e. $ \frac{4}{\gamma} $ corresponds to  10.4 ps. Thus, our model predicts the sufficient increase of coherence lifetime, which was observed experimentally. Although our prediction could overestimate the increase as we assume an ideal resonance between reservoir modes and the transition and as we consider an oversimplified model for the FMO complex. Let us note that the condition for the transition to the oscillatory regime is held $ \gamma^2 - 16 g^2 < 0 $. Moreover, our model predicts the frequency of population oscillations $ |\Delta| \approx 29 \, \text{cm}^{-1} $. The presence of such oscillations does not coincide with the Markovian models in the case of absence of degeneracy but was observed experimentally \cite{Engel07}. Let us emphasize that the presence of degeneracy also allows one to explain such oscillations by the presence of the dark states \cite{Kozyrev18, Kozyrev18a}.

Let us also note that the data of two-dimensional echo spectroscopy were actually observed in the experiment \cite{Mukamel99, Cho09}, namely the dependence of diagonal peaks and cross-peaks on so-called population time was observed. But if we consider four diagrams \cite[p. 178]{Cho09} making contribution to the cross-peaks, then two of them turn to be zero in the one-exciton approximation, one of them makes population-time-independent contribution at zero temperature. And the last one is just defined by coherences dynamics. Two diagrams contribute to diagonal peaks, one of them is also  population-time-independent at zero temperature and the other is defined by the population dynamics. Thus, the relation between the spectroscopy data and reduced density matrix dynamics is direct in our case (for example it is not the case for \cite{Plenio13}).

The pseudomode approach uses the low-temperature approximation. As $  n = \frac{1}{e^{\beta \omega_0}-1} $ contains only the product of the reverse temperature and the frequency, then it could be used for the description of high-frequency peaks considered in \cite{Kolli12}. However, in the next section we offer a possible way to remove this restriction.

\section{Finite temperature case and Markov chain deformation}\label{sec:finTemp}

Proposition \ref{prop:finTempMarkLim} from section \ref{sec:NonMark} allows one two consider non-Markovianity as a ''deformation'' of the initial Markov chain changing the decay to the interaction with a mediating level and a subsequent decay. An important feature of such a deformation is that the initial Markovian dynamics for the reduced density matrix is reproduced as $ \lambda \rightarrow 0 $. So let us extend such an idea to the finite-temperature case. Namely, let us consider the evolution with the generator
\begin{equation}\label{eq:evrGen}
\gamma_0 n D_{01} + \frac{4 g^2}{\gamma_0 (n+1)} D_{\tilde{1} 0} + g h_{1 \tilde{1}}.
\end{equation}
In many aspects this generator is similar to the generator from \cite{Kozyrev17} for the three-level system interacting with several reservoirs and a coherent field.

The next proposition generalizes proposition \ref{prop:finTempMarkLim} guaranteeing transition to the Markovian equation without Lamb and Stark shifts for the reduced density matrix (see \cite[p. 154]{Breuer10}, \cite{Accardi97}) as $ \lambda \rightarrow 0 $.

\begin{proposition}
	Let $ \rho(t, g , \gamma_0, n) $ be a solution of the Cauchy problem
	\begin{equation*}
	\frac{d}{dt} \rho(t, g , \gamma_0, n) = \gamma_0 n D_{01}(\rho(t, g , \gamma_0, n)) + \frac{4 g^2}{\gamma_0 (n+1)} D_{\tilde{1} 0} (\rho(t, g , \gamma_0, n)) + g h_{1 \tilde{1}}(\rho(t, g , \gamma_0, n)).
	\end{equation*}
	Then $ \rho_M(t, \gamma_0, n) \equiv \lim\limits_{\lambda \rightarrow 0} \rho_S \left(\frac{t}{\lambda^2}, \lambda g , \lambda^2 \gamma_0, n \right) $ satisfies the equation
	\begin{equation*}
	\frac{d}{dt} \rho_M(t, \gamma_0, n) = \gamma_0 (n+1) D_{1 0} (\rho_M(t, \gamma_0, n)) + \gamma_0 n D_{01} (\rho_M(t, \gamma_0, n)).
	\end{equation*} 
\end{proposition}
The proof the this proposition is similar to the proof of proposition \ref{prop:finTempMarkLim}. But the expression for $ \rho_S(t, g , \gamma_0, n)  $ (a fortiori for $ \rho(t, g , \gamma_0, n)  $) is much more complex, so we show here only the expression for
\begin{align*}
\rho_M(t, \gamma_0, n) \equiv \lim\limits_{\lambda \rightarrow 0} \rho_S \left(\frac{t}{\lambda^2}, \lambda g , \lambda^2 \gamma_0, n \right) = \left(\frac{n}{1+2n} + \left(\rho_{11}-\frac{n}{1+2n}\right)e^{- \gamma_0 (2n+1) t}\right)| 1 \rangle \langle 1| +\\
+ \left(\frac{1+n}{1+2n} - \left(\rho_{11}-\frac{n}{1+2n}\right)e^{- \gamma_0 (2n+1) t}\right)| 0 \rangle \langle 0| + e^{- \frac{\gamma_0}{2} (2n+1)t} (\rho_{01} | 0 \rangle \langle 1| + \rho_{10} | 1 \rangle \langle 0|),
\end{align*} 
which corresponds to the solution of the Markovian equation.

This proposition allows one to use generator \eqref{eq:evrGen} for description of the non-Markovian phe\-no\-me\-na in the light harvesting complexes at finite (in particular, room) temperature.

\section{Conclusion}
The non-Markovian resonance decay at zero temperature was described by means of the pseudo\-mode approach. The obtained results explain experimentally observed phenomena in the FMO complexes. Unlike \cite{Plenio13} we use much simpler equations which at the same time allow one to examine the role of non-Markovian phenomena in arising long-living coherences and population oscillations in a non-degenerate system. The deformation approach was suggested. It allows one to obtain equations at finite temperature.

In our opinion, the following directions for the further studies should be mentioned:
\begin{enumerate}
	\item To study the non-Markovian phenomena in more detailed models of the FMO complex  as well as other light harvesting complexes at cryogenic temperatures and at room temperatures but for high frequency peaks in spectral density.
	\item To apply the heuristic equations from section \ref{sec:finTemp} for description of the non-Markovian phenomena at room temperature.
	\item To compare the results of calculations by the heuristic equations and by other methods (time-convolutionless projection operator method, hierarchical equations of motion etc.) for the simplest systems.
	\item To expand the theory of Markov chain deformation, in particular, to study how the deformation of one transition in a quantum Markov chain influences the deformation of other transitions.
	\item To attempt to justify the deformation approach strictly beyond zero temperature approximation.
\end{enumerate}

Thus, the results of this work could serve as a basis for the further studies.

\section{Acknowledgments}
The author thanks I.\,V.~Volovich, S.\,V.~Kozyrev, A.\,I.~Mikhailov and A.\,S.~Trushechkin for fruitful discussion of the problems considered in the work.

\end{document}